\documentclass[%
superscriptaddress,
 amsmath,amssymb,
 aps,
 pre,
 twocolumn,
 numerical,
]{revtex4-2}

\usepackage[utf8]{inputenc} 
\usepackage[T1]{fontenc}    
\usepackage{hyperref}       
\usepackage{url}            
\usepackage{booktabs}       
\usepackage{amsfonts}       
\usepackage{nicefrac}       
\usepackage{microtype}      
\usepackage{xcolor}         
\usepackage{amsmath,amsfonts,amssymb,bm}
\usepackage{color}
\usepackage{graphicx}
\usepackage{lmodern}
\usepackage{float}
\usepackage{scalerel}

\begin{document}

\title{Unsupervised and probabilistic learning with Contrastive Local Learning Networks: The Restricted Kirchhoff Machine}

\author{Marcelo Guzman}
\thanks{These authors contributed equally. Email: mguzmanj@sas.upenn.edu, simoneciarella@gmail.com}
 \affiliation{Department of Physics and Astronomy, University of Pennsylvania, Philadelphia PA 19104
}

\author{Simone Ciarella}
\thanks{These authors contributed equally. Email: mguzmanj@sas.upenn.edu, simoneciarella@gmail.com}
 \affiliation{Netherlands eScience Center, Amsterdam, The Netherlands} 
 \affiliation{Laboratoire de Physique de l'Ecole Normale Sup\'erieure, ENS, Universit\'e PSL, CNRS, Sorbonne Universit\'e, Universit\'e de Paris, F-75005 Paris, France
}

\author{Andrea J. Liu}
 \affiliation{Department of Physics and Astronomy, University of Pennsylvania, Philadelphia PA 19104
}
\affiliation{Santa Fe Institute, 
Santa Fe, NM 87501}
\date{\today}

\begin{abstract}
Autonomous physical learning systems modify their internal parameters and solve computational tasks without relying on external computation.
Compared to traditional computers, they enjoy distributed and energy-efficient learning due to their physical dynamics.
In this paper, we introduce a self-learning resistor network, the Restricted Kirchhoff Machine, capable of solving unsupervised learning tasks akin to the Restricted Boltzmann Machine algorithm.
The circuit relies on existing technology based on Contrastive Local Learning Networks, in which two identical networks compare different physical states to implement a contrastive local learning rule.
We simulate the training of the machine on the binarized MNIST dataset, providing a proof of concept of its learning capabilities.
Finally, we compare the scaling behavior of the time, power, and energy consumed per operation as more nodes are included in the machine to their Restricted Boltzmann Machine counterpart operated on CPU and GPU platforms.
\end{abstract}

\maketitle

\section{Introduction} 
Physical dynamics and local interactions have been exploited to implement distributed autonomous  learning~\cite{dillavou2022demonstration,dillavou2024machine, kaiser2022hardware, niu2024self, schneider1993analog}. As opposed to machine learning implemented in a computer, these physical systems learn to perform supervised tasks on their own, without a processor or external memory, presenting advantages in speed and energy consumption~\cite{momeni2024training}.
One class of experimental systems that is potentially scalable to large platforms is electronic Contrastive Local Learning Networks (CLLNs)~\cite{dillavou2022demonstration,dillavou2024machine}.
These electrical networks have edges consisting of self-adjusting linear~\cite{dillavou2022demonstration} or nonlinear~\cite{dillavou2024machine} resistors. When voltages are applied at certain nodes as boundary conditions, each edge dynamically updates its conductance according to a local learning rule~\cite{stern2021supervised,kendall2020trainingendtoendanalogneural} based on the response of the edge to different boundary conditions. Learning of a desired voltage response at output nodes---such as for classification or regression---is distributed throughout the system and emerges as a collective property.
Moreover, the local rule is easily modified in the hardware, implementing several forms of supervised learning including synchronous~\cite{dillavou2022demonstration}, asynchronous~\cite{wycoff2022desynchronous}, non-adiabatic~\cite{stern2022physical}, energy-efficient~\cite{stern2024training}, and robust~\cite{dillavou2025understanding} training.

Here we go beyond autonomous \emph{supervised} learning~\cite{scellier17, kendall2020trainingendtoendanalogneural,dillavou2022demonstration, dillavou2024machine, kaiser2022hardware, stern2021supervised, scellier17,prezioso2015training} and processor-based training of physical systems~\cite{wright2022deep, laydevant2024training, dorband2015boltzmann, niazi2024training, momeni2024training}, to propose a local rule and network design for autonomous \emph{probabilistic} learning, encompassing both unsupervised and supervised settings.
We introduce, in theory and simulations, a variant of CLLNs inspired by the Restricted Boltzmann Machine (RBM)~\cite{ackley1985learning,hinton12} that we call the Restricted Kirchoff Machine (RKM).
The RKM differs from previous CLLNs in incorporating noise, rectifiers, and the bipartite architecture used in RBMs. Like earlier CLLNs, the RKM uses Kirchhoff's law to transform the action of local rules into desired collective behavior, achieving performance on the binarized MNIST dataset that is comparable to RBMs while offering significant scaling advantages in energy and time with increasing system size.

\begin{figure*}[t]
  \centering    \includegraphics[width=\textwidth]{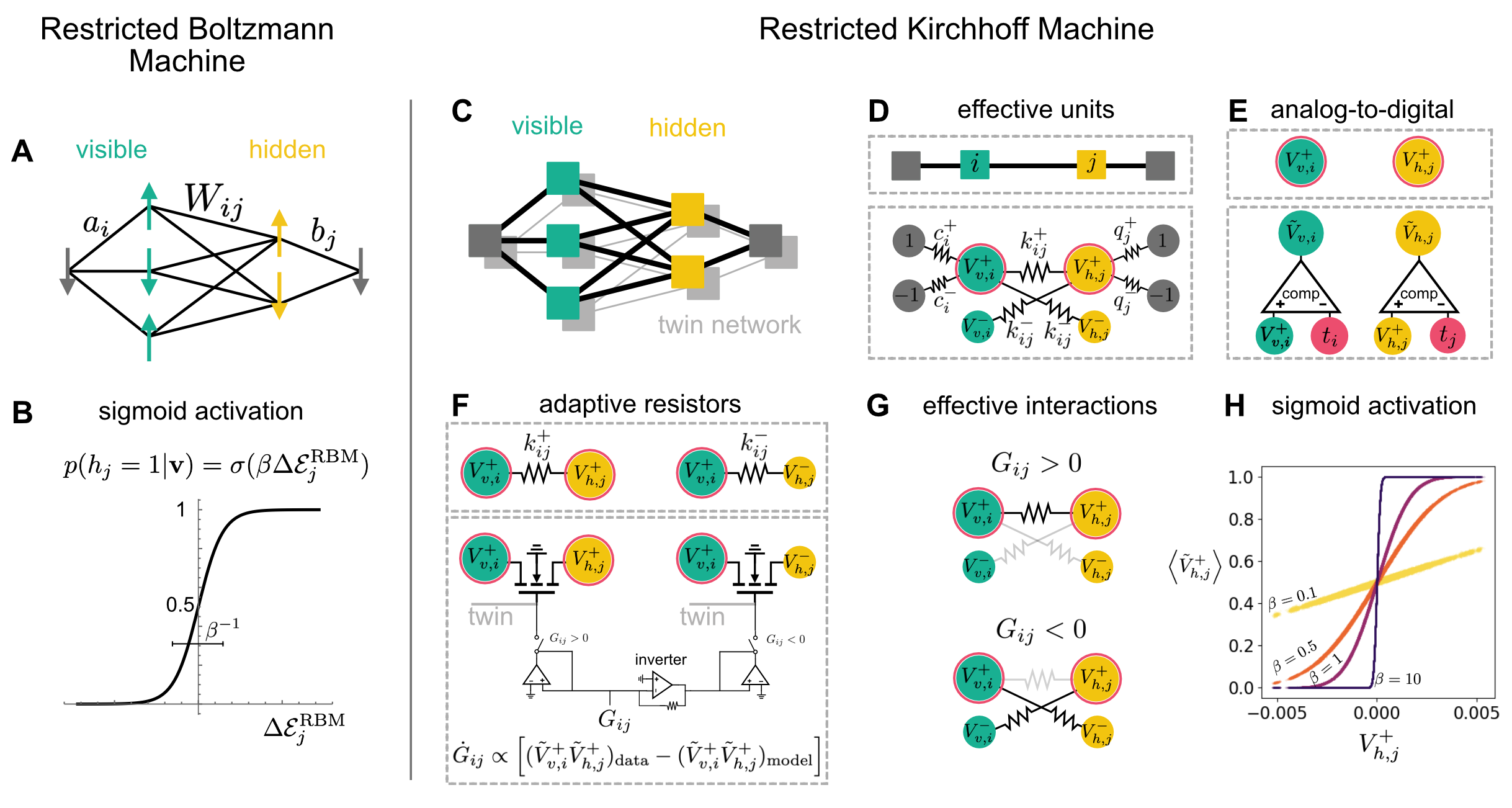}
  \caption{{\bf RBM vs RKM}.
  {\bf A.} The RBM is a binary spin glass system with visible (green), hidden (yellow), and fixed (gray) spins and tunable interactions $W_{ij}$, $a_i$, and $b_j$.
  {\bf B.} The probability of finding spin $j$ up corresponds to the Boltzmann factor of the energy difference $\Delta \mathcal E_j^{\text{RBM}}$.
  {\bf C.} The RKM is composed of two identical networks (color and light gray) with the bipartite architecture of the RBM.
  Each network has visible (green), hidden (yellow), and ground (dark gray) units illustrated by squares.
  {\bf D.} Each unit node corresponds to two analog voltage nodes with positive and negative values (disks), and the interactions between units are mediated by linear adaptive resistors.
 {\bf E.} The binary voltage nodes (red-bordered circles also shown in {\bf  D}) are digital signals corresponding to the output of comparators.
   The comparator inputs are the analog steady voltage value and a random voltage $t$ that changes in time.
   Its output is either 1 if the steady voltage is larger than $t$, or 0 if smaller.
  {\bf F.} The adaptive resistors have conductances controlled by gate voltages $G_{ij}$ following the local learning rule~eq.~\ref{eq:learning} as a function of the terminal values.
  The sign of $G_{ij}$ defines whether it activates the left or right circuit through a combination of operational amplifiers,  inverters, and switches.
  {\bf G.} Effectively, the sign of $G_{ij}$ defines which conductances are non-zero.
  {\bf H.} Effective sigmoid activation for a network of $N_v=784$ and $N_h=500$ units.
  Each curve corresponds to a different $\beta$, and each point to a different pair $(V^+_{h,j},\left<\tilde V^+_{h,j}\right>)$, where the brackets indicate noise average over 10000 realizations.}
  \label{fig:1}
\end{figure*}

\subsection{Summary of Restricted Boltzmann Machine (RBM)}

For a more detailed review of RBMs see~\cite{mehta2019high}.
A Boltzmann Machine is an Ising model with learnable couplings $W_{ij}$ between spins, whose equilibrium Boltzmann distribution is trained to reproduce the statistics of a data set.
A Restricted Boltzmann Machine has the two-layer architecture of Fig.~\ref{fig:1}A: a visible layer with spins $\mathbf v=(v_1,v_2,...)$ encoding the data and a hidden layer with spins $\mathbf h=(h_1,h_2,...)$ serving as latent variables.
The energy of the system is given by
\begin{equation}
    \mathcal E^\text{RBM}(\mathbf v,\mathbf h) =-\sum_{ij}v_iW_{ij}h_j -\sum_iv_ia_i-\sum_jh_jb_j,
    \label{eq.energyRBM}
\end{equation}
where $W_{ij}$ are inter-layer couplings, and $a_i$ and $b_j$ are couplings to frozen spins.
The equilibrium joint distribution is $p(\mathbf v,\mathbf h) =\exp(-\beta \mathcal E^\text{RBM})/Z$, with inverse temperature $\beta$ and partition function $Z$.
The aim is for the model distribution, obtained by marginalizing (integrating) over $\mathbf h$,
\begin{equation}
    p_{model}(\mathbf v)=\sum_\mathbf hp(\mathbf v,\mathbf h)=\sum_\mathbf h\frac{\exp(-\beta \mathcal E^\text{RBM})}{Z},
\end{equation}
to approximate the data distribution $p_{data}(\mathbf v)$.
This can be achieved by minimizing the Kullback-Leibler divergence between the two distributions,
\begin{equation}
D=\sum_{\mathbf v} p_{data}(\mathbf v)\log \frac{p_{data}(\mathbf v)}{p_{model}(\mathbf v)}, \label{eq:KLdiv}
\end{equation}
with $D \ge 0$ and $D=0$ when $p_{data}(\mathbf v)=p_{model}(\mathbf v)$.
The updates to $W_{ij}$, for example, are given by
\begin{equation}
\Delta W_{ij} \propto -\frac{\partial D}{\partial W_{ij}} \propto - \biggl[ \left< \frac{\partial \mathcal E^{\text{RBM}}}{\partial W_{ij}} \right>_{data}-\left< \frac{\partial \mathcal E^{\text{RBM}}}{\partial W_{ij}} \right>_{model} \biggr],
\label{eq:updates}
\end{equation}
leading to
\begin{equation}
    \Delta W_{ij} \propto \left<v_ih_j\right>_{\text{data}}-\left<v_ih_j\right>_{\text{model}},
\label{eq:RBMupdates}
\end{equation}
where the two averages are taken over the data and the model distribution, respectively, with analogous updates for $a_i$ and $b_j$.
In practice, these averages are taken as sample averages, which can be easily computed thanks to the bipartite architecture of the RBM.
The absence of visible-to-visible and hidden-to-hidden connections renders the conditional probabilities independent:

\begin{align}
    p(\mathbf v| \mathbf h)&=\prod_i p(v_i|\mathbf h),\\
    p(\mathbf h| \mathbf v)&=\prod_j p(h_j|\mathbf v),
\end{align}
where each probability is determined by the logistic function $\sigma$ applied to the energy difference (Fig.~\ref{fig:1}B),  e.g.  $p(h_j =1 | \mathbf v) = \sigma (\beta \Delta \mathcal E^{\text{RBM}}_j)$, with $\Delta \mathcal E^{\text{RBM}}_j= \mathcal E^{\text{RBM}}(h_j=1,\mathbf v)- \mathcal E^{\text{RBM}}(h_j=-1,\mathbf v)$, and $\beta$ the inverse temperature.
It follows that the average over data samples $\mathbf v=\mathbf v_{data}$ reduces to sampling from $p(\mathbf h|\mathbf v)$ (forward process).
The average over the model involves  $p(\mathbf v,\mathbf h)$, which can be computed using Gibbs sampling by iteratively applying the forward and backward process: drawing samples $\mathbf h_{t+1}$ from $p(\mathbf h|\mathbf v_t)$ and $\mathbf v_{t+1}$ from $p(\mathbf v|\mathbf h_{t+1})$ until reaching equilibrium~\cite{mehta2019high}.
This is, however, very slow, and different approximations based on truncating the Markov Chain exist, such as contrastive divergence~\cite{hinton2002training} and persistent contrastive divergence~\cite{tieleman2008training}.

\section{The Restricted Kirchhoff Machine}

We now describe the setup for an electrical circuit realization of the RBM algorithm.
Notice that the form of eq.~\ref{eq:updates} is very similar to the conductance updates in experimentally implemented electrical Contrastive Local Learning Networks~\cite{dillavou2022demonstration,dillavou2024machine}.
In particular, in linear resistor networks~\cite{dillavou2022demonstration}, the conductances of the resistor edges are adjusted according to the rule:
\begin{equation}
\Delta k_{ij} \propto - \biggl [ \frac{\partial P_{clamped}}{\partial k_{ij}}-\frac{\partial P_{free}}{\partial k_{ij}} \biggr ],  \label{eq:CL}
\end{equation}
where $P$ is the power dissipated by the system.
Carrying out these derivatives leads to a local rule for updating the conductance of the edge connecting nodes $i$ and $j$ that depends only on the voltages on nodes $i$ and $j$, just as the update rule for the RBM depends only on the spin values $v_i$ and $h_j$ on nodes $i$ and $j$.

Similarly, the conductance update in eq.~\ref{eq:CL} depends on a difference between two states, corresponding to the ``clamped" and ``free" boundary conditions, respectively.
To apply them simultaneously, CLLNs use  twin networks~\cite{dillavou2022demonstration}, in which the conductance of each edge is the same for the networks subjected to the clamped and free boundary conditions.
Here we also propose to use two identical copies of the same network, Fig.~\ref{fig:1}C (color and light gray), to develop electronic realizations of the RBM, which we call the Restricted Kirchhoff Machine (RKM).

Each of the twins contains two layers of analog voltage nodes---visible (green) and hidden (yellow)---and ground nodes (dark gray).
Each spin is encoded in two voltage nodes, so that which of the two nodes has the higher voltage dictates whether the spin is up or down.
We denote by $V_{v,i}^{\pm}$ the two voltages corresponding to spin $i$ in the visible layer, and $V_{h,j}^{\pm}$ the two voltages representing spin $j$ in the hidden layer.
Couplings between spins $i$ and $j$ are encoded in three adjustable resistors connecting the two nodes for each spin, Fig.~\ref{fig:1}D.
This is necessary because conductances are necessarily positive, while the couplings must be allowed to be positive or negative.

Following Ref.~\cite{dillavou2024machine}, the resistors are transistors, whose conductance is controlled by a gate voltage $G_{ij}$.
Each set of three resistors connecting two spins is controlled by the same gate voltage (Fig.~\ref{fig:1}F).
Ferromagnetic interactions are implemented via a single resistor with conductance $k_{ij}^+(G_{ij})$, while antiferromagnetic interactions use two identical resistors with conductances $k_{ij}^-(G_{ij})$.
The same gate voltage controls the three resistors in the twin network, guaranteeing that it is a twin.
The sign of the interaction is determined by the sign of the gate voltage relative to ground and is implemented via a voltage comparator, see Fig.~\ref{fig:1}F and G.
For simplicity, we consider ideal transistors and comparators, for which we have $k^+_{ij}=s G_{ij}\Theta(G_{ij})$ and  $k^-_{ij}= -sG_{ij}\Theta(-G_{ij})$, with $s$ a dimensional prefactor in units of $\text{Ohms}^{-1}\text{Volt}^{-1}$ and assumed equal to 1.

Likewise, we incorporate the fixed spin to the left of the visible layer and the fixed spin to the right of the hidden layer (Figs.~\ref{fig:1}A and C) by applying fixed voltage values of $\pm 1$ to the two nodes representing each of the fixed spins.
The couplings $a_i$ and $b_i$ are captured by resistors with conductances $c_i^{\pm}$ and $q_j^{\pm}$, respectively.
In total, each of the two circuits, of  $N_v$ and $N_h$ effective units, is composed of $2(N_v+N_h+2)$ voltage nodes and $3N_vN_h+2(N_v+N_h)$ adjustable resistors.

The RKM responds to voltages applied to nodes in the visible or hidden layers (boundary conditions) by redistributing currents according to Kirchhoff's law.
In linear resistor networks, this physical response is equivalent to the minimization of electrical power $\mathcal P$~\cite{vadlamani2020physics}.
As we show next, the idea behind the RKM is to replace the energy (Ising Hamiltonian) by the electrical power $\mathcal P$.
To clarify the correspondence, it is convenient to define the difference and sum of conductances as follows:
\begin{equation}
\begin{aligned}
    W_{ij} & = 2(k_{ij}^+ - k_{ij}^-),  & \quad M_{ij} & = 2(k_{ij}^+ + k_{ij}^-), \\
    a_i & = 2(c_i^+ - c_i^-),  & \quad r_i & = 2(c_i^+ + c_i^-), \\
    b_j & = 2(q_j^+ - q_j^-),  & \quad s_j & = 2(q_j^+ + q_j^-),
\end{aligned}
\label{eq:change_var}
\end{equation}
Specifically, in the mapping between the RBM and RKM, the RBM weights are represented by conductance differences in the RKM.

The power dissipated by the RKM is the sum of a convex (C) and non-convex (NC) landscape as a function of the node voltages (see MM~\ref{MM:powerdissipation} for an extensive derivation):
\begin{equation}
    \mathcal P = \mathcal P^{\text{NC}}+\mathcal P^{\text{C}},
    \label{eq:powpow}
\end{equation}
where

\begin{align}
\mathcal P ^{\text{NC}}&=
-\sum_{ij}V_{v,i}^+W_{ij}V_{h,j}^+-\sum_iV_{v,i}^+a_i-\sum_jV_{h,j}^+b_j,\label{eq:pnc}\\
    \mathcal P ^{\text{C}}&=\frac{1}{2}\sum_{ij}M_{ij}\left((V_{v,i}^+)^2+(V_{h,j}^+)^2\right)\nonumber\\
    &\;\;\;\;\;\;+\frac{1}{2}\sum_i (V_{v,i}^+)^2 r_i+\frac{1}{2}\sum_j (V_{h,j}^+)^2s_j.
    \label{eq:pc}
\end{align}
Importantly, the non-convex landscape has the same mathematical form as the Ising spin energy of an RBM in eq.~\ref{eq.energyRBM}, $\mathcal P ^{\text{NC}}=\mathcal E^{\text{RBM}}$.
When visible voltages are clamped to fixed values, with $V_{v,i}^+=-V_{v,i}^-$,  the hidden voltages relax towards the equilibrium values given by
\begin{equation}
V_{h,j}^+=\frac{\sum_iV_{v,i}^+W_{ij}+b_j}{\sum_iM_{ij}+s_j}=\frac{\Delta \mathcal E_j^{\text{RBM}}}{\sum_iM_{ij}+s_j}.
    \label{eq:hiddeneq}
\end{equation}
Notice that the final expression is proportional to the spin energy difference of the RBM model with hidden spin $j$ in the up and down states.

\begin{figure*}[t]
  \centering
\includegraphics[width=\textwidth]{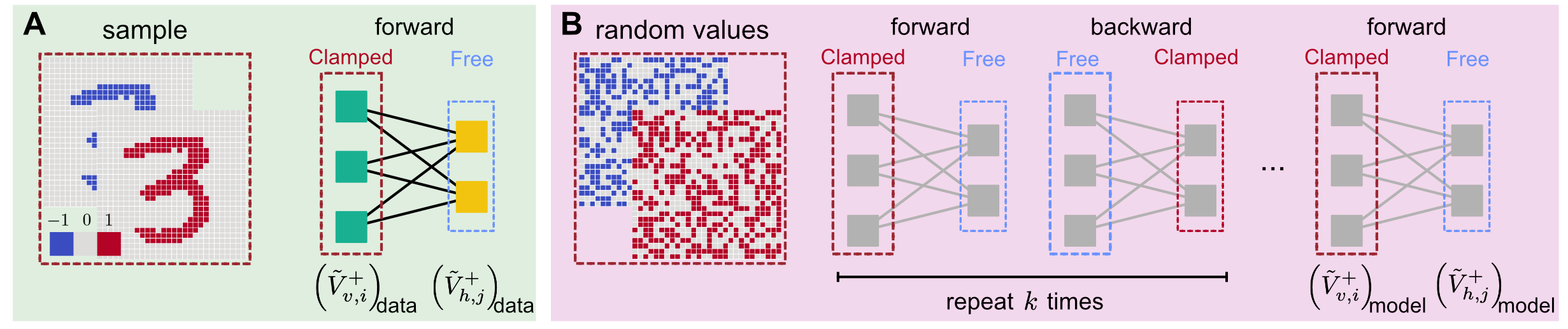}
  \caption{{\bf Contrastive learning in the RKM.}
  {\bf A.} One network processes samples of the training data by clamping voltages to the visible layer and performing a forward passage. The final configuration corresponds to the $D$ state.
  {\bf B.} The twin network initially processes a random voltage configuration on its visible layer, which is then propagated forward and backward $k$ times. The final configuration corresponds to the $R$ state.
  Once both networks are equilibrated, the self-adapting edges locally compute~eq.~\ref{eq:learning} and update the gate voltage $G_{ij}$ accordingly.
  The final visible state of the twin network is then reused as the initial state for the next learning iteration.}
  \label{fig:learning}
\end{figure*}

There are still two very important differences between spins in the RBM and corresponding voltage node pairs in the RKM.
First, the spins are discrete while voltage differences take continuous values. Second, in the RBM, the energy difference $\Delta \mathcal E_j^{\text{RBM}}$ determines the probability of the spin being in the up state through the Boltzmann factor (see Fig.~\ref{fig:1}B). In the RKM,  the equilibrium configuration is deterministically decided by eq.~\ref{eq:hiddeneq}.
To binarize the voltages and add stochasticity we use a further trick.
We compare each equilibrated analog voltage with a random voltage $t_j$:
\begin{equation}
    \tilde V_{h,j}^+=\Theta(V_{h,j}^+-t_j)\in \{0,1\},
    \label{eq:binary}
\end{equation}
where $\Theta$ is the Heaviside function.
This analog-to-digital transformation is local and implemented by comparators as shown in Fig.~\ref{fig:1}E.
If the random voltage $t_j$ is chosen from a Gaussian distribution with zero mean and variance $(\beta N_v)^{-1}$, the analog-to-digital operation of eq.~\ref{eq:binary} on average implements a sigmoidal activation function, see Fig.~\ref{fig:1}H and MM~\ref{MM:noise}. Equations~eq.~\ref{eq:hiddeneq} and ~eq.~\ref{eq:binary} thus constitute a {\it forward} passage of the RKM.
Notice that the notion of temperature is controlled by the amplitude of the random signal $t_j$.

We highlight that this analog-to-digital transformation is not identical to the sigmoid function applied to the energy difference, as in an RBM.
Indeed, to isolate $\Delta\mathcal E_j^{\text{RBM}}$ in eq.~\ref{eq:hiddeneq}, we would have to multiply $V_{h,j}^+$ by the denominator $\sum_i M_{ij}+s_j$, which is non-local, as it entails all the visible nodes through the sum over $i$.

Equivalently, in the {\it backward} passage,  the hidden voltages are clamped to constant values $V_{h,j}^+=-V_{h,j}^-$, the visible voltages relax and then are binarized obtaining:

\begin{align}
    V_{v,i}^+&=\frac{\sum_jW_{ij}V_{h,j}^++a_i}{\sum_jM_{ij}+r_i}=\frac{\Delta \mathcal E_i^{\text{RBM}}}{\sum_jM_{ij}+r_i},
    \label{eq:visibleeq}\\
    \tilde V_{v,i}^+&=\Theta(V_{v,i}^+-t_i)\in \{0,1\},\label{eq:binary2}
\end{align}
with $t_i\sim\mathcal N(0,\frac{1}{\beta N_h})$.


\subsection{Contrastive Learning in the RKM}
The learning rule of the RKM corresponds to eq.~\ref{eq:RBMupdates} approximated by Persistent Contrastive Divergence (PCD)~\cite{tieleman2008training, tieleman09} and physically implemented by two copies of the same network with internal feedback loops as in Ref.~\cite{dillavou2024machine}.

At each learning iteration, the two networks simultaneously relax toward different states, the data and the model state.
In the first network, a training sample is clamped to the visible layer and the hidden configuration is obtained from a forward passage, with the joint configuration constituting a data state, Fig.~\ref{fig:learning}A.
In the twin network, the model state is computed by PCD-k, in which an initially random visible configuration is clamped, followed by $k$ forward and backward processes, see Fig.~\ref{fig:learning}B.
The last visible configuration is then used as the starting point for the next iteration in the twin network, playing the role of the persistent chain~\cite{tieleman2008training}.
The gate voltages are then updated as  $G_{ij} \rightarrow G_{ij}+\Delta G_{ij}$ with
\begin{equation}
    \Delta G_{ij}= \alpha\left[ (\tilde V_{v,i}^+\tilde V_{h,j}^+)_{data}-(\tilde V_{v,i}^+\tilde V_{h,j}^+)_{model}-2\lambda G_{ij}\right],
    \label{eq:learning}
\end{equation}
where $\alpha$ is the learning rate, and $\lambda>0$  implements a ridge (or L2) regularization, locally penalizing large values of $G_{ij}$.
Similarly, the gate voltages $G_i$ for conductances $c_i^{\pm}$, and $G_j$ for conductances $q_j^{\pm}$ are updated by

\begin{align}
    \Delta G_i &= \alpha \left[(\tilde V_{v,i}^+)_{data}-(\tilde V_{v,i}^+)_{model}-2\lambda G_i\right],\\
    \Delta G_j &= \alpha \left[(\tilde V_{h,j}^+)_{data}-(\tilde V_{h,j}^+)_{model}-2\lambda G_j\right].
\end{align}

Note that in RBMs, eq.~\ref{eq:RBMupdates} involves ensemble averages; this requires memory storage. To avoid using external memory the RKM uses the contrast of single states (online learning).
After training, only one of the twin networks is needed to perform the inference.

\section{Numerical results}
As a proof of concept, we simulate the training of the RKM on the black and white, handwritten digit images of the MNIST dataset.
Each sample of the training data corresponds to an image of $N_v=28\times28$ pixels. This sets the size of the visible layer, with two nodes in each of the twin networks specifying the ``spin" state of each pixel (white or black).
We use $N_h=500$ hidden units, and simulate the training for $3\times10^7$ learning iterations, learning rate $\alpha = 0.0001$, $\lambda = 0.001$, and standard initialization tricks detailed in the Supplementary Information~\ref{MM:hyperpar}.

\begin{figure}[t]
  \centering
\includegraphics[width=\columnwidth]{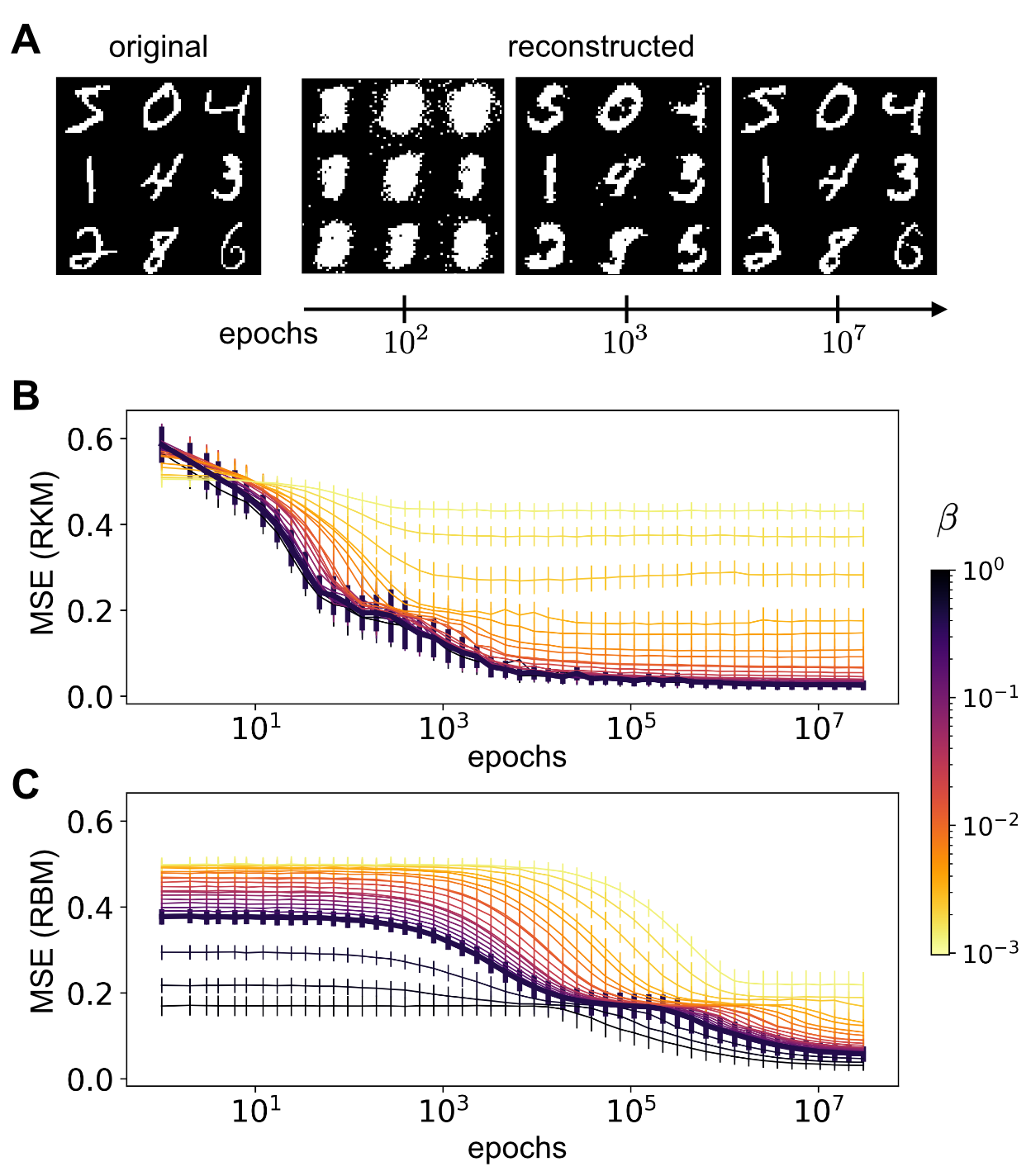}
  \caption{{\bf Reconstruction performance.}
  {\bf A.} Examples of reconstructed images at different training epochs for $\beta=0.05$.
  {\bf B.} Evolution of the MSE over 500 test samples as a function of the training epochs for different $\beta$ (color).
  The thickest line corresponds to $\beta=0.12$, leading to the best final MSE.
  {\bf C.} Same as {\bf D} for an RBM with identical architecture and hyperparameters as the RKM.
  As before, the thickest line corresponds to $\beta=0.12$, however, it does not lead to the best reconstruction performance.}
  \label{fig:reconstruction}
\end{figure}

We evaluate the RKM’s learning performance through reconstruction and generation. In reconstruction, we measure how well the model retains test samples by comparing each image in the test set to its forward-backward version, see Fig.~\ref{fig:reconstruction}A.
We quantify the reconstruction error using the mean squared error (MSE) over all pixels and test samples:
\begin{equation}    \text{MSE}=\sum_n^{N_{\text{samples}}}\sum_{\nu}^{N_{\text{pixels}}}(p^{\text{original}}_{n,\nu}-p^{\text{reconstructed}}_{n,\nu})^2.
\end{equation}

The evolution of the MSE along the training of the RKM is plotted in Fig.~\ref{fig:reconstruction}B for different values of $\beta$.
For all cases, training reduces the reconstruction MSE, with the best performance observed for $\beta>0.04$.
Notice that the performance improves negligibly past that point, and the error plateaus.
For comparison, an RBM trained with the identical hyperparameters and architecture achieves similar performance, as shown in Fig.~\ref{fig:reconstruction}C.
Here, however, there is no saturation, and the larger the $\beta$, the better the reconstruction.

\begin{figure}[t]
  \centering
\includegraphics[width=\columnwidth]{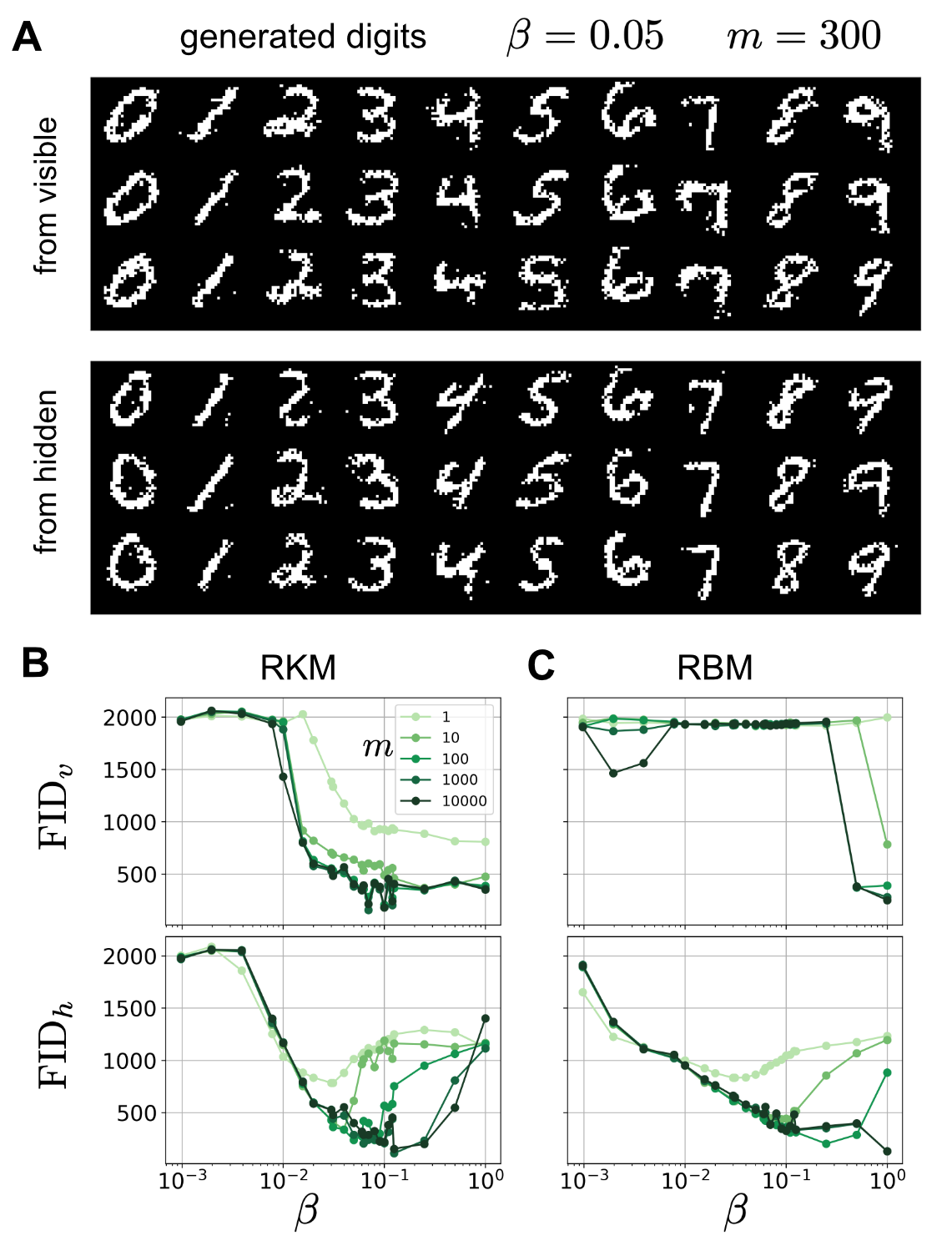}
  \caption{{\bf Generation performance.}
  {\bf A.} Examples of generated digits with $m=300$ iterations from an RKM trained with $\beta=0.05$.
  The samples were generated from random initial configurations at the visible (top) and hidden (bottom) layers.
  {\bf B.} FID score from random visible (top) and hidden (bottom) states at the end of training as a function of $\beta$ and $m$ generative iterations (color).
  {\bf C.} Same as {\bf B} for an RBM with identical architecture and hyperparameters as the RKM.
  }
  \label{fig:generation}
\end{figure}

In generation, we assess whether the model can produce realistic samples from the learned distribution through the Fréchet Inception Distance (FID) score~\cite{fid} between the test set and a set of generated samples.
The FID is defined as the Fréchet distance between two gaussian distributions, with mean and covariance defined by the samples of the test data  set ($\bm m_{test}$, $C_{test}$) and the generated data set ($\bm m_{gen}$, $C_{gen}$) respectively:
\begin{equation}
    \text{FID}=|\bm m_{test}-\bm m_{gen}|^2+\text{tr}(C_{test}+C_{gen}-2(C_{test}C_{gen})^{1/2}).
\end{equation}
To generate new images, we employ Gibbs sampling~\cite{geman84}, propagating an initially random configuration back and forth $m$ times, clamped at either the visible or hidden layer.
Figure~\ref{fig:generation}A shows samples generated with $m=300$ iterations from an RKM trained with $\beta=0.05$, starting with random configurations at the visible (top) and hidden (bottom) layers.
Unlike standard RBM sampling~\cite{tieleman2008training}, this method avoids the use of memory to store a persistent chain~\cite{decelle21}, making it more suitable for the RKM.

Figure~\ref{fig:generation}B summarizes the FID score of the RKM for $k=1$ and different choices of  $\beta$  and the number of iterations in the generation process, $m$.
Unlike the reconstruction task, noise is required for generative capability; in the deterministic case, the network tends to memorize training samples and cannot generate new ones, leading to a high FID score, regardless of the starting layer.
Instead, we find that the FID score has a non-monotonic behavior with respect to both $\beta$ and $m$, with $\beta\sim 0.05-0.3$ being optimal for generation.
In addition, we find that generating samples from initially random values in the hidden layer generically leads to lower FID scores.
For comparison, we also show the FID score performance for an RBM trained under the same conditions, following the same generation scheme as the RKM, see Fig.~\ref{fig:generation}C.
The results demonstrate the RKM's ability to learn similarly to an RBM but in a decentralized hardware instead of digital software using a centralized computer.

\begin{figure}
\begin{center}
\includegraphics[width=\columnwidth]{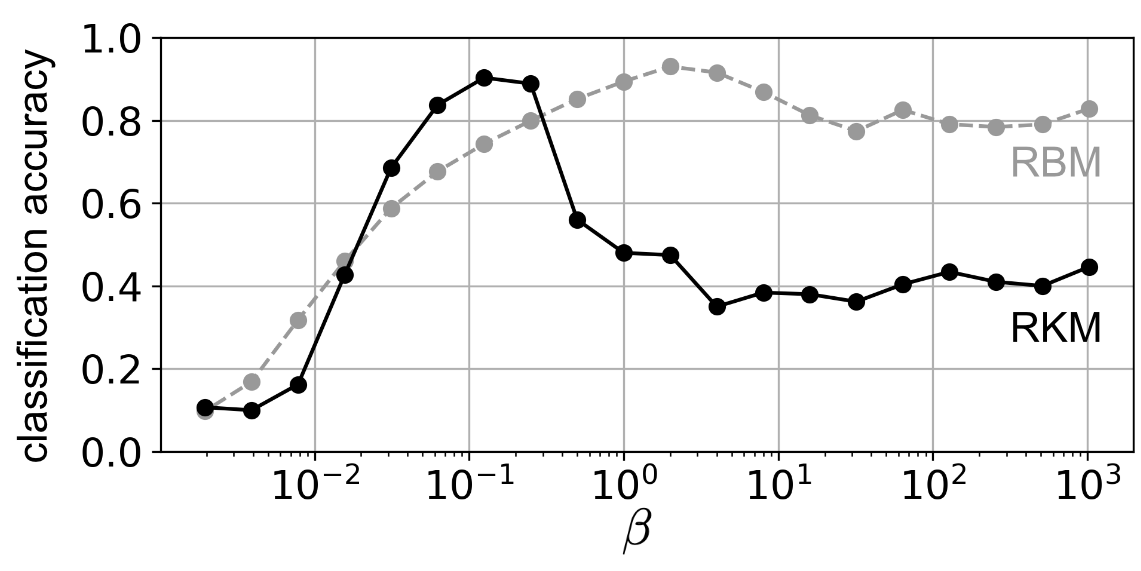}
\caption{{\bf Classification performance.}
Classification accuracy over test data (500 samples) for RKMs (black line) trained at different temperatures and $k=1$.
To classify an image, only the first 10 visible inputs, corresponding to the one-hot encoding of the label, are updated in the back-and-forth passes, while the remaining inputs are kept fixed.
Each point represents the average accuracy from $m=300$ to $m=500$  back-and-forth iterations.
Error bars indicating standard deviation are smaller than the point size.
For comparison, the gray line shows the accuracy of  RBMs trained under the same data and hyperparameters.
Both systems reach the same peak accuracy, but the RKM plateaus at a lower value than the RBM for large $\beta$.}
\label{fig:classification}
\end{center}
\end{figure}

The unsupervised learning algorithm for RBMs can also be used for classification, by training a model that generates both the data and the labels~\cite{larochelle2012learning,hinton2006fast}.
This can be achieved, for example, by including additional inputs as one-hot encoding of the labels.
To test the performance of RKMs on MNIST classification, we include 10 additional inputs to represent digit labels, so that along with the image, a voltage input of 1 is applied to the correct label while voltage inputs of 0 are applied to the other 9 inputs. Again, we use no batching or memory.
The results are shown in Fig.~\ref{fig:classification} as a function of inverse temperature $\beta$; as expected, the test accuracy (over 500 test samples) is non-monotonic in $\beta$ with a peak at $\beta \approx 0.1$ at around $92 \%$.
For comparison, we also show the test accuracy of an RBM trained under the same conditions (gray line in Fig.~\ref{fig:classification}) revealing a comparable peak performance, with optimal $\beta$ shifted to larger values.
In addition, for larger $\beta$ values, the RBM  plateaus at an accuracy of $80\%$, while the RKM does so at $40\%$.
This difference stems from the nature of the binarization and noise injection.
At high $\beta$, both the sigmoid activation of the RBM and the analog-to-digital conversion of the RKM become step functions.
However, the arguments of these step functions are different: for the RBM it corresponds to the energy difference $\Delta\mathcal E_j^\text{RBM}$, while for the RKM it is the normalized energy difference, $\Delta\mathcal E_j^\text{RBM}/(\sum_iM_{ij}+s_j)$, see eq.~\ref{eq:hiddeneq}.

This proof of principle shows that minimal physical components can support probabilistic learning for unsupervised and supervised setups. The performances of the RKM on reconstruction, generation, and classification are comparable to an RBM under similar constraints (without batching or memory).
For the RKM trained on the binarized MNIST dataset, we find that the optimal temperature range is $\beta\sim 0.04-0.3$, consistent across all the tasks considered.
We note that it is possible for the RKM to go beyond online learning by integrating gate voltages with capacitors~\cite{dillavou2025understanding} in order to compute averages over many training examples and model states.

\section{Scaling relations for time and energy advantage of RKMs over RBMs}

The RKM is inherently distributed and implements local learning in parallel directly in physical components. In other words, it performs ``in-memory learning," avoiding the need for centralized processors and the von Neumann bottleneck~\cite{dillavou2024machine}. Once trained, inference is performed through the physical relaxation to the steady state of the response to input voltages.
Here we analyze the scaling of the time and energy costs of RKMs and compare them to operations required for an RBM on a conventional CPU and GPU.

We start with inference. The inference process of the RKM relies on three core operations per node, representing the main sources of time and energy consumption: analog voltage relaxation (A), noise generation (N), and analog-to-digital conversion (AD).
While the time, power, and energy required to perform each of these operations will depend on the specifics of the electronic components, we can nevertheless describe how they scale with system size.


First, we focus on operations N and AD, enforced by the comparators shown in Fig.~\ref{fig:1}E.
We notice that they are distributed across the nodes and independent of the network state.
It follows that the time to perform any of the two operations does not depend on the number of units $N_v + N_h$ (the number of nodes being $2(N_v+N_h)$), while the power per operation does so linearly.
The energy consumed by these operations, on the other hand, does not simply depend on how long it takes to perform the operations individually, but on the total time needed for the inference process, as the comparators must be active all the time.
Mathematically, if $T_o$, $P_o$, and $E_o$ are the time, power, and energy required for operation $o$, and $\tau$ is the total time for an inference process, we have $E_o = P_o\times \tau$, with $\tau \geq T_{\text{N}}+T_{\text{AD}}+T_{\text{A}}$ to ensure that all operations are satisfied.






We next focus on the analog relaxation time (A).
Information propagates through the network by the analog voltage response set by Kirchhoff's current law.
The time to reach the steady state will depend on the (parasitic) capacitances of the circuit~\cite{dillavou2022demonstration,dillavou2024machine} and will be different for the forward $\text{A}_\text{F}$ and backward processes $\text{A}_\text{B}$.
We model such an effect as capacitors with constant capacitance $C$ connecting each voltage node to the ground and interpret the mean RC time per node as the time required per operation.
Due to the bipartite nature of the network, each node in the visible layer is connected in parallel to every node in the hidden layer and vice versa, leading to (see Supplementary Information \ref{MM:noise} for details):

\begin{align}
    &T_{A_F}(N_v,N_h) = 2C\frac{1}{N_h}\sum_j^{N_h} \left(\sum_i^{N_v}M_{ij}+s_j\right)^{-1}\sim \mathcal O (N_v^{-1}),
    \label{eq:time_h}\\
    &T_{A_B}(N_v,N_h) =2C\frac{1}{N_v}\sum_i^{N_v} \left(\sum_j^{N_h}M_{ij}+r_i\right)^{-1}\sim \mathcal O (N_h^{-1})
    \label{eq:time_v}.
\end{align}
Since the RC time decreases with the number of parallel connections, we find that the forward (backward) time is inversely proportional the number of visible (hidden) units---the operation gets faster as the size goes up.

In the learning phase, there is an additional operation per learning iteration to consider, associated with updating the gate voltages (G). As with N and AD, this operation is local and independent of the network state, leading to a time independent of the number of tunable gate voltages (edges). The power scales linearly with the number of tunable edges ($3N_v N_h+2(N_v+N_h$). Note that since the time associated with updating gate voltages $T_{\text G}$ is independent of $N_v$ and $N_h$, the time required for an inference or learning step scales in the same way for the RKM.

\begin{figure}
\begin{center}
\includegraphics[width=\columnwidth]{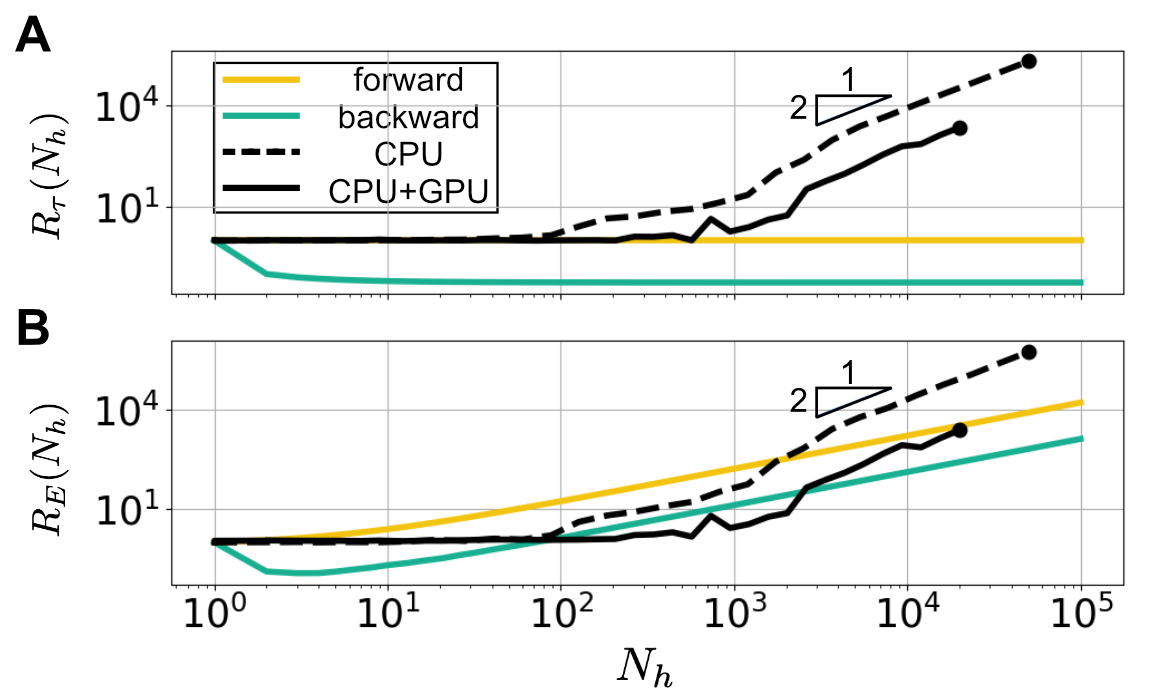}
\caption{{\bf Time and energy scaling relations}.
Time ratio ({\bf A}) and energy ratio ({\bf B}) as functions of the number of hidden nodes for the forward and backward process in the RKM (color) and matrix-vector multiplication in a centralized computer (black).
For the RKM, brackets indicate averages over 200 gate voltage values $G_{ij}$ and 200 input voltages.
For the matrix-vector multiplication, $10^5$ randomly initialized \textit{torch.matmul} operations are averaged for each point.
The black disk indicates the largest size that we could process on our hardware consisting of an Intel i7-13700H and an NVIDIA 4050 GPU.}
\label{fig:4}
\end{center}
\end{figure}

It follows from these arguments that the overall time for a forward or backward process is dominantly independent of $N_v$ and $N_h$: $\tau\sim\mathcal O(1)$.
This is numerically confirmed in Fig.~\ref{fig:4}A, where we consider $\tau = T_{\text{N}}+T_{\text{AD}}+T_{\text{A}}$ and we plot the time ratio
\begin{equation}
R_{\tau}(N_h)\equiv\frac{\left<\tau(N_v=784,N_h)\right>}{\left<\tau(N_v=784,1)\right>}
\end{equation}
as a function of $N_h$, with the brackets denoting the average over 200 different sets of initial conductances and 200 random input voltages. The number of visible units is set by the sample size of the MNIST dataset.
For comparison, we show the time ratio required to perform a matrix-vector multiplication of a weight matrix of dimension $N_v\times N_h$---a fundamental operation in the RBM algorithm---as a function of $N_h$.
For both CPU and GPU-accelerated processes, the time ratio grows faster than linearly with $N_h$ holding $N_v=784$ fixed.

\begin{table*}[t]
\centering
\begin{tabular}{|c ||c|c||c|c||c|c||}
\hline
& \multicolumn{2}{|c||}{Time ($T_o$)}&\multicolumn{2}{|c||}{Power ($P_o$)}& \multicolumn{2}{|c||}{Energy ($E_o=P_o\times \tau$)} \\[0.5ex]
  &{\small Forward} & Backward  &Forward & Backward  &Forward & Backward \\ [0.5ex]
 \hline
 Noise (N) & $\mathcal O(1)$ & $\mathcal O(1)$ & $\mathcal O(N_h)$ & $\mathcal O(N_v)$ & $\mathcal O(N_h)$ & $\mathcal O(N_v)$\\
 Analog-to-Digital (AD) & $\mathcal O(1)$ & $\mathcal O(1)$ & $\mathcal O(N_h)$ & $\mathcal O(N_v)$ & $\mathcal O(N_h)$ & $\mathcal O(N_v)$\\
  Analog relaxation (A) & $\mathcal O(N_v^{-1})$ & $\mathcal O(N_h^{-1})$& $\mathcal O(N_vN_h)$ & $\mathcal O(N_vN_h)$& $\mathcal O(N_vN_h)$ & $\mathcal O(N_vN_h)$\\[0.5ex]
\hline
{\bf Total (inference)} &$\mathcal O(1)$ &$\mathcal O(1)$&$\mathcal O(N_vN_h)$&$\mathcal O(N_vN_h)$&$\mathcal O(N_vN_h)$&$\mathcal O(N_vN_h)$\\
 \hline
    \cline{2-7}
Gate voltage update (G) &  \multicolumn{2}{|c||}{$\mathcal O(1)$}& \multicolumn{2}{|c||}{$\mathcal O(N_vN_h)$}&\multicolumn{2}{|c||}{$\mathcal O(N_vN_h)$}\\
\hline
{\bf Total (inference+update)} &$\mathcal O(1)$ &$\mathcal O(1)$&$\mathcal O(N_vN_h)$&$\mathcal O(N_vN_h)$&$\mathcal O(N_vN_h)$&$\mathcal O(N_vN_h)$\\
\hline
\end{tabular}
\caption{\\[0.5ex]
\normalfont
Scaling of time, power, and energy for processes required for inference (N, AD, A) and training (N, AD, A and G) with the numbers of visible and hidden units, $N_v$ and $N_h$, respectively, for the Restricted Kirchhoff Machine. Here, $N$ corresponds to noise generation, $AD$ to analog-digital conversion, $A$ to analog voltage relaxation, $G$ to gate voltage updates.}
\label{table:1}
\end{table*}

Our argument for $\tau\sim\mathcal O(1)$ implies that both power and energy have the same scaling per operation in the RKM.
For N and AD, they become extensive with the total number of relevant nodes: $P_o\sim E_o\sim \mathcal O(N_h)$ for the forward and $P_o\sim E_o\sim \mathcal O(N_v)$ for the backward process.
While less evident, the same scaling dominates the operation A, for which the power dissipated is mainly the power $\mathcal P$ of  eq.~\ref{eq:powpow} evaluated at the steady voltage configurations  eq.~\ref{eq:hiddeneq} (forward) and eq.~\ref{eq:visibleeq} (backward).
Fig.~\ref{fig:4}B shows the total energy ratio $R_{E}(N_h) \equiv\left<E(N_v,N_h)\right>/\left<E(N_v,1)\right>$, where $E=E_A+E_N+E_{AD}$, for the forward and backward process, which increases linearly with $N_h$.
For comparison, the energy consumption of the matrix-vector multiplication, with both CPU and GPU acceleration, approaches a quadratic scaling with $N_h$.
The scaling relations are summarized in table~\ref{table:1}.
Finally, in the SI we show how the MSE and FIDs scores vary with the system size,  confirming the expected trend of improved performance with more hidden nodes.

\section{Discussion}
We have introduced the Restricted Kirchhoff Machine (RKM), a variant of contrastive local learning networks (CLLNs) capable of performing unsupervised or supervised learning and generative tasks.
By leveraging the principles of Kirchhoff's laws and the architecture of Restricted Boltzmann Machines (RBMs), the RKM demonstrates the potential for decentralized and scalable probabilistic learning in resistor networks.

Our in-silico training over the binarized MNIST dataset demonstrates the RKM's ability to learn. We find that the RKM approaches the performance of an RBM for reconstruction, generative and classification tasks for the same hyperparameters.
The presence of noise, rectifiers, and a bipartite structure enables the RKM to learn complex data distributions through local learning rules without relying on external memory.
Importantly, it builds upon the existing technology of CLLNs, for which tunable resistors implementing local rules are already in use.

Due to the parallel and distributed nature of the RKM, we find that regardless of the electronic details, the scaling of time, power, and energy consumption for training is better than for RBMs on centralized platforms by a factor of $1/N_h$, when the number of visible units $N_v$ is kept fixed.

From a practical perspective, an experimental RKM can benefit from several optimization protocols not included here, where we have considered only the minimal ingredients.
In particular, the learning performance could be significantly improved by batching, where the network averages the learning rule~eq.~\ref{eq:learning} over several samples before modifying the gate voltages.
Such an approach can be implemented by capacitors storing the successive updates when rapidly switching between samples~\cite{dillavou2025understanding}.
In addition, there are several directions to optimize the physical efficiency of the machine: implementing an optimal layout to minimize the parasitic capacitance $C$, updating gate voltages before reaching a steady voltage configuration~\cite{stern2022physical}, using small gate voltage values to reduce the conductances and power dissipation, and using correlated noise to lower the number of noise generators and thus decrease $P_N$, to name a few.
Moreover, as in previous work, it is straightforward to modify the local learning rule to penalize high-power dissipation solutions and find low-power ones ~\cite{stern2024training}.
In general, we expect these two metrics---learning performance and physical performance---to present trade-offs that will be analyzed in future work.

The RKM belongs to a family of systems known as Ising machines, where computers, algorithms, and physical systems are used to find the ground state of an Ising model to \textit{e.g.}, solve combinatorial problems~\cite{mohseni2022ising,niazi2024training}.
In the context of physical learning, most of these approaches rely on hybrid set-ups, combining centralized computers with physical systems such as field programmable gate arrays~\cite{niazi2024training, patel2022logically}, D-wave systems~\cite{laydevant2024training,dorband2015boltzmann}, memristors~\cite{aguirre2024hardware}, and in general, a wide variety of physical systems~\cite{wright2022deep}.
The RKM, on the other hand, implements both inference and training on the same physical substrate.
Compared to other self-adapting Ising models, such as those based on magnetic tunnel junctions~\cite{kaiser2022hardware}, spintronics~\cite{niu2024self}, or atomic interactions~\cite{kiraly2021atomic}, the CMOS-based tunable resistor design that we propose here offers a promising path to microfabrication and scalability~\cite{dillavou2022demonstration,dillavou2024machine,dillavou2025understanding}.

More broadly, this work takes a step toward more capable autonomous physical learning platforms. Unlike previous systems~\cite{stern2023learning}, the RKM blends analog and digital signals in its learning process, reminiscent of biological neural computation~\cite{debanne2013mechanisms}, opening the field of physical learning up to the study of hybrid analog-digital architectures. From a theoretical perspective, it opens new avenues for exploring unsupervised learning dynamics in resistor networks and understanding how they navigate both error and power landscapes~\cite{stern2025physical, guzman2025microscopic}.

\emph{Acknowledgements.}---This work was primarily supported by the NSF through DMR-MT-2005749 (MG, AJL). The project was initiated at an annual meeting of the Simons Collaboration on ``Cracking the Glass Problem" via Award \# 348126; Additional support from the Simons Foundation was provided via Glass Collaboration Award \# 454955 (SC) and Investigator Award\# 327939 (AJL).  AJL is also grateful for the hospitality of the Center for Computational Biology at the Flatiron Institute, the Newton Institute for Mathematical Sciences at Cambridge University (EPSRC grant EP/R014601/1), the Aspen Center for Physics (NSF grant PHY-2210452) and the Santa Fe Institute.

\emph{Data availability.}--- The data and the software supporting these findings are openly available on GitHub~\cite{pyrkm}.

\bibliography{biblio}

\section{Supplementary Information}

\subsection{Power dissipation\label{MM:powerdissipation}}
Each edge connecting nodes $V_{v,i}^+$ and $V_{h,j}^+$ in the network carries a current proportional to the voltage difference and its conductance, $I_{ij}^+=k_{ij}^+\left(V_{v,i}^+-V_{h,j}^+\right)$, and dissipates a power $I_{ij}^+\left(V_{v,i}^+-V_{h,j}^+\right)$.
A linear resistor network with the connectivity described in the main text dissipates a total power given by

\begin{align}
\mathcal P_{\text{total}}
   &= \sum_{ij}\Bigl[
        k_{ij}^+\,(V_{v,i}^+ - V_{h,j}^+)^2 \notag\\
   &\qquad\quad
        + k_{ij}^-\,(V_{v,i}^+ - V_{h,j}^-)^2
        + k_{ij}^-\,(V_{v,i}^- - V_{h,j}^+)^2
      \Bigr] \notag\\
   &\quad + \sum_i\Bigl[
        q_i^+(1 - V_{v,i}^+)^2
        + q_i^-(1 - V_{v,i}^-)^2
      \Bigr] \notag\\
   &\quad + \sum_j\Bigl[
        c_j^+(1 - V_{h,j}^+)^2
        + c_j^-(1 - V_{h,j}^-)^2
      \Bigr].
\end{align}
where the sums are over the $N_v$ and $N_h$ effective units.
We next apply the change of variables indicated in the main text:
\begin{equation}
\begin{aligned}
    W_{ij} & = 2(k_{ij}^+ - k_{ij}^-),  & \quad M_{ij} & = 2(k_{ij}^+ + k_{ij}^-), \\
    a_i & = 2(c_i^+ - c_i^-),  & \quad r_i & = 2(c_i^+ + c_i^-), \\
    b_j & = 2(q_j^+ - q_j^-),  & \quad s_j & = 2(q_j^+ + q_j^-),
    \label{eq:changeofvariables}
\end{aligned}
\end{equation}
leading to
\begin{equation}
\mathcal P_{\text{total}}=\mathcal P^{NC}+\mathcal P^{C}+\mathcal P_{\text{res}}+ \mathcal P_{0},
\label{eq:SIpower}
\end{equation}
where

\begin{align}
\mathcal P^{NC}
   &= -\sum_{ij}V_{v,i}^+W_{ij}V_{h,j}^+
      - \sum_iV_{v,i}^+a_i
      - \sum_jV_{h,j}^+b_j \\[1ex]
\mathcal P^{C}
   &= \tfrac{1}{2}\sum_{ij}M_{ij}\Bigl[(V_{v,i}^+)^2 + (V_{h,j}^+)^2\Bigr] \notag\\
   &\quad + \tfrac{1}{2}\sum_i (V_{v,i}^+)^2 r_i
      + \tfrac{1}{2}\sum_j (V_{h,j}^+)^2 s_j \\[1ex]
\mathcal P_{\text{res}}
   &= \tfrac{1}{2}\sum_ir_i
      + \tfrac{1}{2}\sum_js_j \\[1ex]
\mathcal P_{0}
   &= \tfrac{1}{4}\sum_{ij}(M_{ij} - W_{ij})\Bigl[
        (V_{v,i}^-)^2 + (V_{h,j}^-)^2 - 2V_{v,i}^+V_{h,j}^+ \notag\\
   &\qquad - 2V_{v,i}^-V_{h,j}^+ - 2V_{v,i}^+V_{h,j}^-
      \Bigr].
\end{align}
$\mathcal P_{res}$ is the residual power dissipation due to the presence of ground voltages, independent of the visible and hidden voltage values, and $\mathcal P_0$ determines the state of the negative voltage units, $V_{v,i}^-$ and $V_{h,j}^-$.
Since the RKM always has clamped values in one of the layers, the steady-state voltage values $V_{v,i}^+$ and $V_{h,j}^+$ are only determined by the first two contributions: $\mathcal P =\mathcal P^{\text{NC}}+\mathcal P ^{\text{C}}$.
To demonstrate this statement, let us compute the hidden configuration in the forward passage.
Here, $V_{v,i}^+=-V_{v,i}^-$ and $\mathcal P_0$ is independent of the hidden configuration $V_{h,j}^+$, thus, by taking the gradient of the total power dissipated, we have:

\begin{align}
    \left.\frac{\partial \mathcal P_{\text{total}}}{\partial V_{h,j}^+}\right|&_{\left\{V_{v,i}^+=-V_{v,i}^-\right\}}=\left.\frac{\partial \mathcal P}{\partial V_{h,j}^+}\right|_{\left\{V_{v,i}^+=-V_{v,i}^-\right\}}\nonumber\\&=-\left(\sum_iV_{v,i}^+W_{ij}+b_j\right)+V_{h,j}^+\left(\sum_iM_{ij}+s_j\right),
\end{align}
and imposing the gradient to be zero, we end up with the voltage response in the hidden layer
\begin{equation}
V_{h,j}^+=\frac{\sum_iV_{v,i}^+W_{ij}+b_j}{\sum_iM_{ij}+s_j}.
\label{eq:SIeq1}
\end{equation}
Following the same calculations, we can compute the state of the negative nodes, given by
\begin{equation}
    V_{h,j}^-=\frac{\sum_i V_{v,i}^+(M_{ij}-W_{ij})}{\sum_i (M_{ij}-W_{ij})}.
    \label{eq:SIeq2}
\end{equation}
For the backward passage, in which $V_{h,j}^+=-V_{h,j}^-$, the same calculations lead to

\begin{align}
&V_{v,i}^+=\frac{\sum_jW_{ij}V_{h,j}^++a_i}{\sum_jM_{ij}+r_i},
    \label{eq:SIeq3}\\
    &V_{v,i}^-=\frac{\sum_j V_{h,j}^+(M_{ij}-W_{ij})}{\sum_j (M_{ij}-W_{ij})}.
    \label{eq:SIeq4}
\end{align}

\subsection{Noise and rectifier\label{MM:noise}}
As described in the main text, the digitized voltage response $\tilde V_{h,j}^+$ comes from a comparison of the analog voltage with a normally distributed random variable $t_j\sim \mathcal N(0,(\beta N_v)^{-1})$:
\begin{equation}
\tilde V_{h,j}^+=\Theta(V_{h,j}^+-t_j)=\Theta\left(\frac{\sum_iV_{v,i}^+W_{ij}+b_j}{\sum_iM_{ij}+s_j}-t_j\right),
\end{equation}
where $\Theta$ is the Heaviside function.
Since $t_j$ is normally distributed, the probability of having $\tilde V_{h,j}^+=1$ is
\begin{align}
p\left(\tilde V_{h,j}^+=1| t_j\right)&=p\left(V_{h,j}^+>t_j\right)\nonumber\\&=\frac{1}{2}\left[1-\text{Erf}\left(-\beta N_vV_{h,j}^+/\sqrt{2}\right)\right],
\end{align}
where Erf is the error function, providing the sigmoid activation shown in Fig. 1 of the main text.

Defining  $K\equiv\sum_{ij}M_{ij}/(2N_vN_h)$ as the average conductance of the circuit and denoting by overbars the dimensionless quantities $\bar M_{ij}=M_{ij}/K$ and $\bar s_j=s_j/K$, to first order we have
\begin{equation}
    \sum_i M_{ij}+s_j = K\left(\sum_i \bar M_{ij}+\bar s_j\right)\approx KN_v,
\end{equation}
leading to
\begin{align}
p\left(\tilde V_{h,j}^+=1| t_j\right)&\approx \frac{1}{2}\left[1-\text{Erf}\left[-\frac{\beta}{\sqrt{2}K}\left(\sum_iV_{v,i}^+W_{ij}+b_j\right) \right]\right]\nonumber\\&=\frac{1}{2}\left[1-\text{Erf}\left[-\frac{\beta}{\sqrt{2}K}\Delta \mathcal E_{j}^{\text{RBM}} \right]\right],
\end{align}
i.e., the probability of finding the voltage with value 1 behaves as a sigmoid function on $\Delta \mathcal E_j^{\text{RBM}}=\left(\sum_iV_{v,i}^+W_{ij}+b_j\right)$, similar to the RBM algorithm.

It is worth noting that fewer noise generators may be desirable as they would reduce the energetic cost of adding stochasticity (see section~\ref{sec:scaling}). It remains to be studied how noise correlations among nodes affect the learning capabilities of the RKM.

\subsection{Physical relaxation time}
\begin{figure}
  \centering    \includegraphics[width=0.4\textwidth]{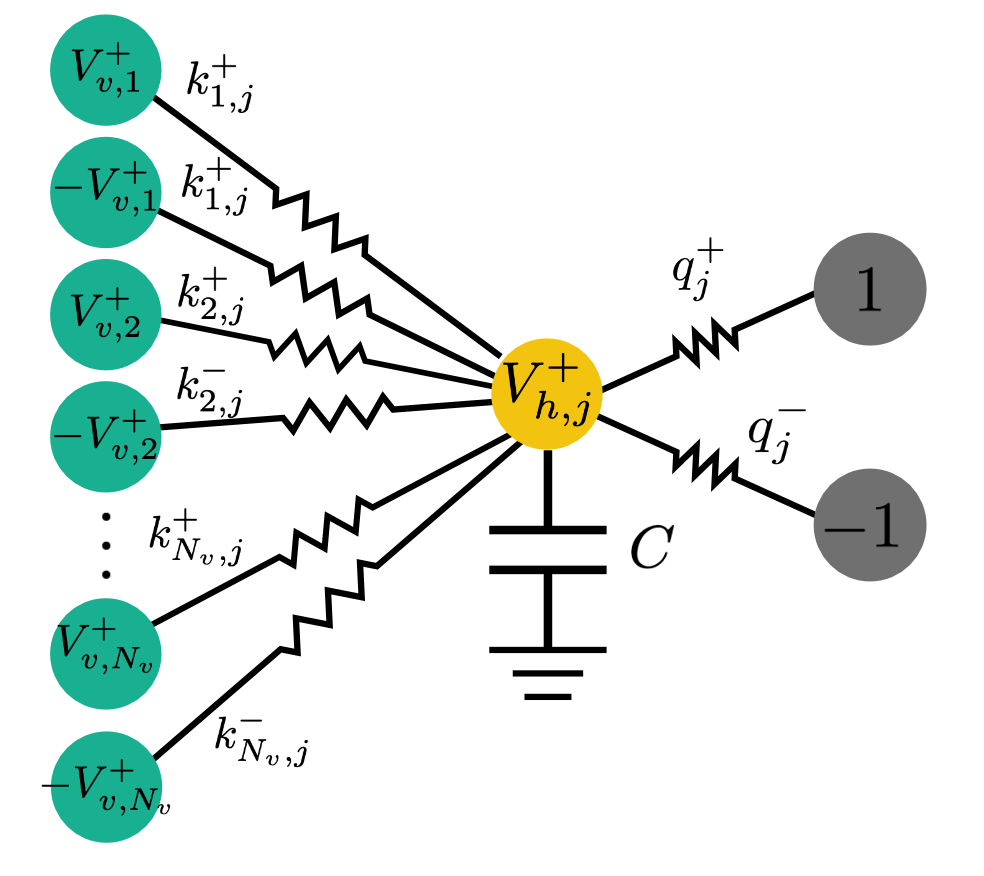}
  \caption{{\bf Effective dynamics in the forward process.} Each voltage node in the hidden layer is attached to $2N_v$ voltage nodes in the visible layer, two voltage nodes $\pm1$ and a capacitor $C$ connected to the ground $V_{\text{ground}}=0$.}
  \label{fig:SI1}
\end{figure}

To compute the relaxation time, we consider a capacitor with capacitance $C$ connected between each voltage node and a ground $V_{ground}=0$ as shown in fig.~\ref{fig:SI1}.
Due to the distributed nature of the circuit, we can analyze the relaxation time of a single node, for example, in the forward passage, where $V_{v,i}^-$ is fixed and $V_{v,i}^-=-V_{v,i}^+$. From current balance at the node $V_{h,j}^+$ we have

\begin{align}
C\frac{d V^+_{h,j}}{dt}
   &= -\Biggl(
        \sum_i k^+_{ij}(V^+_{h,j}-V_{v,i}^+)
        + \sum_i k^-_{ij}(V^+_{h,j}+V_{v,i}^+)
      \Biggr) \notag\\
   &\quad - q^+_j(V_{h,j}^+-1)
          - q^-_j(V_{h,j}^++1) \notag\\
   &= -V_{h,j}^+\tfrac{1}{2}\Bigl(\sum_i M_{ij}+s_j\Bigr)
      + \tfrac{1}{2}\Bigl(\sum_i V_{v,i}^+W_{ij}+b_j\Bigr),
   \label{eq:dynamics}
\end{align}
where in the last line we use the variables defined in~eq.~\ref{eq:changeofvariables}.
The solution of eq.~\ref{eq:dynamics} corresponds to an exponential relaxation towards the equilibrium state:
\begin{align}
V_{h,j}^+(t)&=\left(\frac{\sum_iV_{v,i}^+W_{ij}+b_j}{\sum_iM_{ij}+s_j}\right) \nonumber\\&\quad\quad+\;\;  e^{-t/T_j}\left[V_{h,j}^+(0)-\left(\frac{\sum_iV_{v,i}^+W_{ij}+b_j}{\sum_iM_{ij}+s_j}\right)\right],
\end{align}
with
\begin{equation}
T_j=2C\left(\sum_iM_{ij}+s_j\right)^{-1}.
\end{equation}

Analogously, for the backward passage, we have

\begin{align}
V_{v,i}^+(t)&=\left(\frac{\sum_jV_{h,j}^+W_{ij}+a_i}{\sum_jM_{ij}+r_i}\right) \nonumber\\&\quad+ \;\;e^{-t/T_i}\left[V_{v,i}^+(0)-\left(\frac{\sum_jV_{h,j}^+W_{ij}+a_i}{\sum_jM_{ij}+r_i}\right)\right],
\end{align}
with
\begin{equation}
T_i=2C\left(\sum_jM_{ij}+r_i\right)^{-1}.
\end{equation}

Finally, as done in the main text,  we define the total analog relaxation time for the forward and backward passage as the average of the corresponding nodes:
\begin{align}
    T_{A_F} &= 2C\frac{1}{N_h}\sum_j \left(\sum_iM_{ij}+s_j\right)^{-1},\label{eq:SITAF}\\
    T_{A_B} &= 2C\frac{1}{N_v}\sum_i \left(\sum_jM_{ij}+r_i\right)^{-1}.\label{eq:SITAB}
\end{align}
For convenience, we also define the dimensionless relaxation times:
\begin{equation}
    \bar T_{A_{F/B}}= T_{A_{F/B}}K/C.
    \label{eq:SIdimTAFB}
\end{equation}

\subsection{Scaling analysis of the RKM\label{sec:scaling}}

This section provides a detailed analysis of the scaling relations shown in the main text.
We split the analysis into time, power, and energy per network in the inference phase and report both the forward and backward processes.
Since the data set defines the number of visible units $N_v$, we only refer to numerical results in terms of $N_h$ and consider $N_v=784$.
The numerical results are summarized in Fig.~\ref{fig:SI2}.

\subsubsection*{Time}
We denote by $\tau _F$ and $\tau _B$ the times needed for a forward and backward process:

\begin{align}
\tau_F &= T_{A_F}+ T_{AD_F}+ T_{N_F},\\
\tau_B &= T_{A_B}+ T_{AD_B}+ T_{N_B}
\end{align}
where, following the main text, we consider the analog relaxation (A), the analog-to-digital conversion (AD), and the noise injection (N).
Since AD and N are independent processes, their time of execution is intensive with the number of nodes: $T_{AD_F}+ T_{N_F}=T_{AD_B}+ T_{N_B}= \gamma$, where $\gamma$ is the time to execute both operations in a single voltage node.
We thus have $\tau_{F/B}(N_v,N_h)=T_{A_{F/B}}(N_v,N_h)+\gamma$.

We next study the scaling relations as a function of $N_h$.
To avoid a notation overload, we omit the $N_v$ dependence, but for all cases, it will remain fixed at $N_v=784$ as in the main text.
To provide statistically significant measures, we analyze averages over $n_k=200$ conductance and $n_i=200$ input realizations, denoted by
$\left<\cdot \right>$.
Moreover, to minimize the number of hyperparameters governing the scaling relations, we analyze ratios between properties at size $N_h$ and size $1$.
Then, the ratio of the average time is given by
\begin{align}
    \frac{\left<\tau_{F/B}(N_h)\right>}{\left<\tau_{F/B}(1)\right>} =\frac{\left<\frac{C}{K\gamma}\bar T_{A_{F/B}}(N_h)+1\right>}{\left<\frac{C}{K\gamma}\bar T_{A_{F/B}}(1)+1\right>},
\end{align}
where in the last line we used eq.~\ref{eq:SIdimTAFB}.
The ratio mainly depends on the system size and the dimensionless parameter $\frac{C}{K\gamma}$.
Fig.~\ref{fig:SI2} shows the scaling relations for the backward (first column) and forward (fourth column) process for different values of $\frac{C}{K\gamma}$.
As expected, the forward process is not affected by $N_h$.
In the backward process, however, as the system gets larger, it needs less time to execute the operations.
This effect is magnified as the analog relaxation becomes the slowest process (higher $C/(K\gamma)$).

\subsubsection*{Power}
Since AD and N are independent processes, it follows that their joint power consumption is proportional to the system size.
More precisely, $P_{AD_F}+P_{N_F} = \eta N_h$ and $P_{AD_B}+P_{N_B} = \eta N_v$, with $\eta$ the power consumed by a single unit.
The power dissipated due to the analog relaxation is given by eq.~\ref{eq:SIpower} evaluated at the equilibrated values of ~eq.~\ref{eq:SIeq1} and ~eq.~\ref{eq:SIeq2} (forward), and ~eq.~\ref{eq:SIeq3} and~eq.~\ref{eq:SIeq4} (backward):

\begin{align}
P_{A_F}
   &= \mathcal P_{\text{total}}\Biggl(
        V_{h,j}^+ =
           \frac{\sum_i V_{v,i}^+ W_{ij} + b_j}
                {\sum_i M_{ij} + s_j}, \notag\\
   &\qquad\qquad
        V_{h,j}^- =
           \frac{\sum_i V_{v,i}^+ (M_{ij} - W_{ij})}
                {\sum_i (M_{ij} - W_{ij})}
      \Biggr), \\[1ex]
P_{A_B}
   &= \mathcal P_{\text{total}}\Biggl(
        V_{v,i}^+ =
           \frac{\sum_j W_{ij} V_{h,j}^+ + a_i}
                {\sum_j M_{ij} + r_i}, \notag\\
   &\qquad\qquad
        V_{v,i}^- =
           \frac{\sum_j V_{h,j}^+ (M_{ij} - W_{ij})}
                {\sum_j (M_{ij} - W_{ij})}
      \Biggr).
\end{align}

As before, we decompose the power in terms of an overall scale $KV^2$, where $V$ is the average over the clamped voltages, $P_{A_F}=K V^2\bar P_{A_F}$.
Then, the ratio of the average total power is
\begin{equation}
    \frac{\left<P_F(N_h)\right>}{\left<P_F(1)\right>} =\frac{\left<\frac{K V^2}{\eta}\bar P_{A_F}(N_h)+N_h\right>}{\left<\frac{K V^2}{\eta}\bar P_{A_F}(1)+1\right>},
\end{equation}
determined by the dimensionless parameter $K V^2/\eta$.
Its behavior is shown in the second and fourth columns of Fig.~\ref{fig:SI2}, revealing a weak dependence on the hyperparameter values (color and rows).

\subsubsection*{Energy}
\begin{figure*}
  \centering    \includegraphics[width=\textwidth]{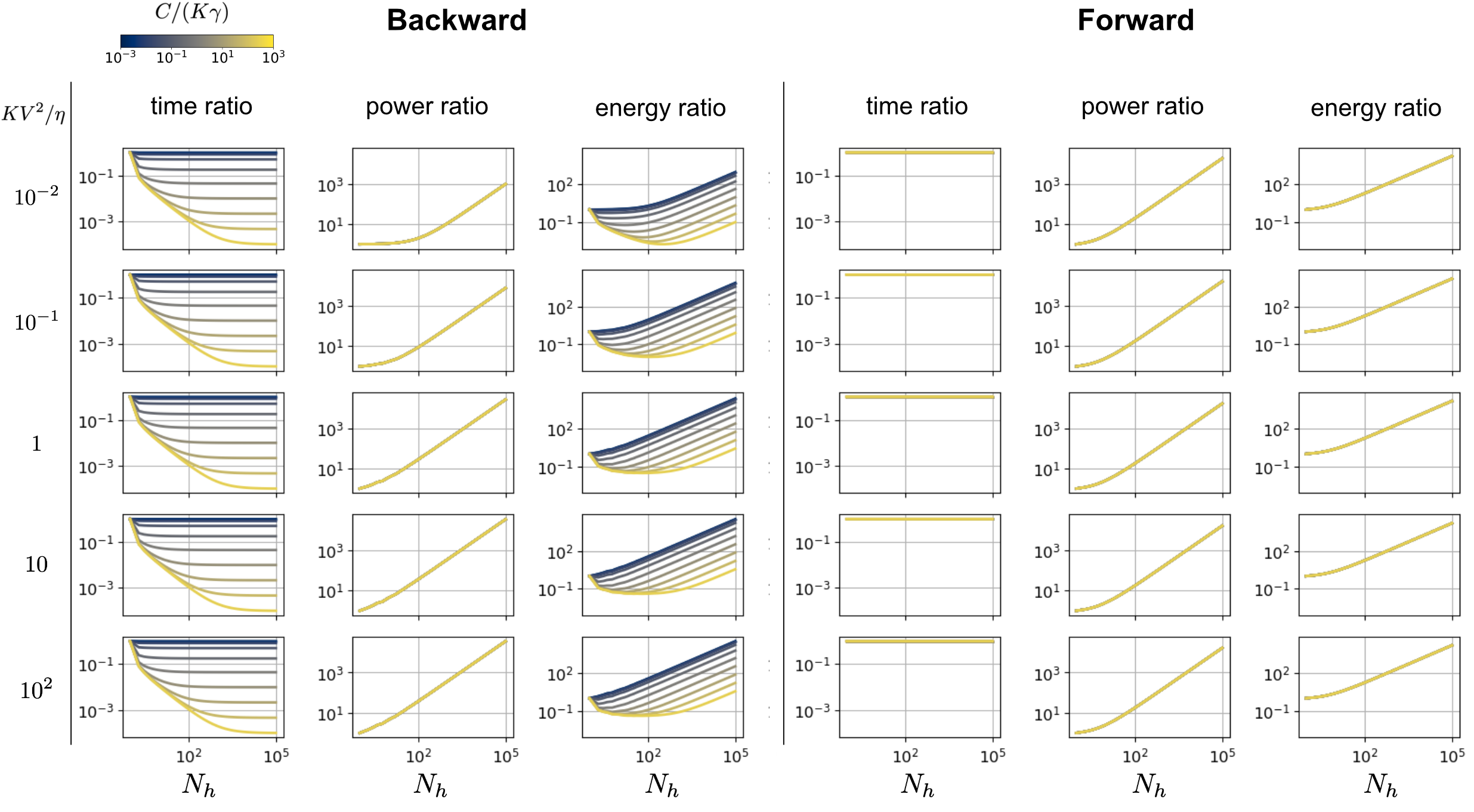}
  \caption{{\bf System size dependence of time, power, and energy consumption} for the backward and forward processes. Colors indicate different values of the dimensionless time parameter, and rows indicate the values of the dimensionless power parameter.}
  \label{fig:SI2}
\end{figure*}

We estimate the energy as the total power times the total time of all the operations:  $E_{F/B} = P_{F/B} \tau_{F/B}$.
The energy ratio depends on the previous two dimensionless parameters,
\begin{align}
    \frac{\left<E_F(N_h)\right>}{\left<E_F(1)\right>}&=\frac{\left<\left(\frac{KV^2}{\eta}\bar P_{A_F}(N_h)+N_h\right)\left(\frac{C}{K\gamma}\bar T_{A_F}(N_h)+1\right)\right>}{\left<\left(\frac{KV^2}{\eta}\bar P_{A_F}(1)+1\right)\left(\frac{C}{K\gamma}\bar T_{A_F}(1)+1\right)\right>},\\
    \frac{\left<E_B(N_h)\right>}{\left<E_B(1)\right>}&=\frac{\left<\left(\frac{KV^2}{\eta}\bar P_{A_B}(N_h)+N_v\right)\left(\frac{C}{K\gamma}\bar T_{A_B}(N_h)+1\right)\right>}{\left<\left(\frac{KV^2}{\eta}\bar P_{A_B}(1)+N_v\right)\left(\frac{C}{K\gamma}\bar T_{A_B}(1)+1\right)\right>}.
\end{align}
The scaling behavior is plotted in the third and sixth columns of Fig.~\ref{fig:SI2}.

\subsection{Scaling analysis of the Matrix-Vector multiplication}

The scaling results reported in the main text for the matrix-vector multiplication have been computed using:
(I) \textit{pyRAPL}: A Python package leveraging Intel's Running Average Power Limit (RAPL) technology to estimate CPU power consumption, and (ii) \textit{nvidia-smi}: The standard NVIDIA System Management Interface to measure GPU energy consumption.

Our code and measured data are publicly available in our GitHub repository~\cite{pyrkm}. The hardware used for measurements includes an Intel i9-9980HK CPU and an NVIDIA GeForce GTX 960M GPU.

\subsection{Training hyperparameters}
\label{MM:hyperpar}
All the machines that we have reported in the main paper follow a series of choices that have been inspired mainly by Ref.~\cite{hinton12} and Ref.~\cite{decelle21} and modified to work for the RKM.
To start, only the first $10000$ images of the binarized MNIST were used (thresholding at the pixel value 0.5). The test set consisted of $5000$ other images. The results reported in the main text use the following hyperparameters
\begin{itemize}
    \item Number of hidden nodes: $N_h=500$.
    \item Learning rate: $\alpha = 0.0001$.
    \item Minibatch size: $n_{mb} = 1$.
    \item Ridge regularization: $\lambda=0.001$.
    \item Initialization: weights $W_{ij}$, corresponding to the gate voltages $G_{ij}$, are normally distributed around zero with standard deviation $0.1/\sqrt{N_v}$.
    The visible biases $a_i$ are initialized as the average values of the training data, and the hidden biases as zeros.
    \item All the gate voltages are clipped into the interval $\in \left[-10,10\right]$ to represent the limited range of variability of the conductances in the RKM circuit.
    \item Persistent Contrastive Divergence training algorithm, where the number of Markov chains used to estimate the negative term of the gradient is $k=n_{mb}=1$.
    \item The number of Gibbs sampling steps used for the negative chains is  $m=1$.
\end{itemize}
All the trainings were done on an Nvidia Titan V GPU, using the PyTorch library.
Our code is publicly available on GitHub~\cite{pyrkm}, and all the results can be fully reproduced following this strategy.

\end{document}